\documentclass{appolb}
\usepackage{graphicx}
\usepackage{amsmath,bm}
\begin{document}
\title{ Gauge-covariant diagonalization of $\pi- a_1$ mixing  and the resolution of a low energy theorem*}
\thanks{Presented at Excited QCD 2020. ($^3$) Speaker. The authors acknowledge support from CFisUC  and FCT through the project UID/FIS/04564/2016 and  grant CERN/FIS-COM/0035/2019, and the networking support by the COST Action CA16201.}
\author{A. A. Osipov($^1$), M. M. Khalifa($^2$), B. Hiller($^3$)
\address{ ($^1$)Bogoliubov Laboratory of Theoretical Physics, Joint Institute for Nuclear Research, Dubna, 141980, Russia;($^2$)  Moscow Institute of Physics and Technology, Dolgoprudny, Moscow Region 141701, Russia and Department of Physics, Al-Azhar University, Cairo 11751, Egypt, ($^3$) CFisUC, Department of Physics, University of Coimbra, P-3004-516 Coimbra, Portugal; } }
\maketitle
\begin{abstract}
Using a recently proposed gauge covariant diagonalization of  $\pi a_1$-mixing we show that the low energy theorem $F^{\pi}=e f_\pi^2 F^{3\pi}$ of current algebra, relating the anomalous  form factor  $F_{\gamma\to\pi^+\pi^0\pi^-}=F^{3\pi}$  and the anomalous neutral pion form factor  $F_{\pi^0\to\gamma\gamma}=F^\pi$, is fulfilled in the framework of the  Nambu-Jona-Lasinio (NJL) model, solving a long standing problem encountered in the extension including vector and axial-vector mesons. At the heart of the solution is the presence of a $\gamma \pi  {\bar q} q $ vertex which is absent in the conventional treatment of diagonalization and leads to a deviation from the vector meson dominance (VMD) picture. It contributes to a gauge invariant anomalous tri-axial  (AAA) vertex as a pure surface term.
\end{abstract}
 \vspace{0.5cm}
 

The Wess-Zumino \cite{Wess71} effective action, with topological content clarified by Witten \cite{Witten83}, describes all effects of QCD anomalies in low-energy processes with photons and Goldstone bosons, without reference to massive vector mesons.  
The extension to the case with spin-1 mesons is not unique, and has been addressed in different frameworks \cite{Schechter84},\cite{Fujiwara85},\cite{Kaiser90}.  
Important issues arise when one includes the spin-1 states. Here we address the concept of VMD and the pseudoscalar -- axial-vector mixing ($\pi a_1$-mixing) of meson states. In particular, it has been shown in \cite{Fujiwara85} that the {\it complete} VMD is not valid in either $\pi^0\to \gamma\gamma$ or $\gamma\to 3\pi$ processes, and that  mixing affects hadronic amplitudes in \cite{Gasiorovicz69,Osipov85}. Therefore one should demonstrate how the departure from VMD occurs and how $\pi a_1$-mixing is treated in order to comply with the predictions of the Wess-Zumino action. This is not a trivial task, in \cite{Wakamatsu89} it has been reported that in  a number of well-known models 
\cite{Schwinger67}-\cite{Bando85c} 
the $\pi a_1$-mixing breaks low-energy theorems (LET) for some anomalous processes, e.g., $\gamma\to 3\pi$, $K^+K^-\to 3\pi$. 
In \cite{okh:2020}, based on the gauge covariant treatment of $\pi a_1$-mixing, only recently addressed \cite{Osipov18a}-\cite{Osipov20}, we show precisely how the deviation of the complete VMD occurs  in the framework of the NJL Lagrangian,  fulfilling the LET 
\begin{equation}
\label{LET}
F^{\pi}=e f_\pi^2 F^{3\pi}.
\end{equation}
 The procedure is sufficiently general to be applied in other processes. 

To be more definite, recall that the $\pi a_1$ diagonalization is generally performed by a linearized transformation of the axial vector field
\begin{equation}
\label{ngcr}
a_\mu \to a_\mu + \frac{\partial_\mu \pi}{ag_\rho f_\pi}, 
\end{equation}
where $\pi =\tau_i \pi^i$, $a_\mu =\tau_i a^i_\mu$ and $\tau_i$ are the $SU(2)$ Pauli matrices; $g_\rho\simeq \sqrt{12\pi}$ is the coupling of the $\rho$ meson to two pions and $f_\pi \simeq 93$MeV the pion  weak decay constant.  In extensions of the model that couple to the electroweak sector this replacement violates gauge invariance \cite{Osipov18a}-\cite{Osipov20} in anomalous processes, leaving however the real part of the action invariant \cite{Osipov18c,Osipov19}. For example the  anomalous low energy amplitude describing the $a_1\to\gamma\pi^+\pi^-$ decay is not transverse \cite{Osipov18a,Osipov18b}.
To restore gauge invariance the gauge covariant derivative $\mathcal D_\mu \pi$ must be used instead of $\partial_\mu \pi$ \cite{Osipov18a}-\cite{Osipov20}  
\begin{equation}
\label{cov}
a_\mu \to a_\mu + \frac{\mathcal D_\mu \pi}{ag_\rho f_\pi}, \quad \mathcal D_\mu \pi =\partial_\mu \pi -ieA_\mu [Q,\pi ], \quad  Q=\frac{1}{2}(\tau_3+\frac{1}{3}),
\end{equation}
In the context of the LET $F^{\pi}=e f_\pi^2 F^{3\pi}$  mixing  occurs related to both anomalous form factors, but it has been proven in \cite{okh:2020} that the radiative decay ${\pi^0\to\gamma\gamma}$ is not affected by the mixing, and coincides with the low energy result of current algebra given by the Lagrangian density \cite{Wess71,Witten83}
\begin{equation}
\label{pigg}
\mathcal{L}_{\pi\gamma\gamma}=-\frac{1}{8}F^\pi\pi^0 e^{\mu\nu\alpha\beta} F_{\mu\nu}F_{\alpha\beta}, \quad F^{\pi}=\frac{N_c e^2}{12\pi^2 f_\pi},
\end{equation}
where $e$ is the electric charge, $F_{\mu\nu}=\partial_\mu A_\nu -\partial_\nu A_\mu$ stands for the strength of the electromagnetic field, $N_c$ is the number of quark colors. The absence of mixing is seen as follows. In the NJL model one can switch to spin-1 variables without direct photon-quark coupling, as described in the VMD picture. Then $\mathcal{L}_{\pi\gamma\gamma}$ is related to the $\pi^0\omega\rho$ quark triangle  shown in Fig. 1(a) left . At leading order of a derivative expansion  the current-algebra result $\Gamma (\pi^0\to\gamma\gamma)=7.1 \, \mbox{eV}$ is obtained. 
Diagram 1 (b) left , due to mixing, is described by an axialvector vector vector (AVV) Adler-Bell-Jackiw anomaly \cite{Adler71}-\cite{Jackiw00}. The related surface term (ST) which results from the difference of two linearly divergent amplitudes is apriori arbitrary. Here this arbitrary parameter is fixed on gauge invariant grounds of $a_1\rightarrow \gamma \gamma$, upon which graph 1 (b) left vanishes at leading order of a derivative expansion. This complies with the Landau-Yang theorem \cite{Landau48},\cite{Yang56} which states that a massive unit spin particle cannot decay into two on shell massless photons.

Effects of $\pi a_1$-mixing in  $\gamma\to 3\pi$ amplitudes (due to G-parity it is sufficient to consider the isoscalar component of the photon, related to $\omega\to 3\pi$) have been studied by Wakamatsu \cite{Wakamatsu89} in detail, using the prescription (\ref{ngcr}). He found that the amplitude of the $\omega\to 3\pi$ decay contains uncompensated contributions generated by $\pi a_1$-mixing, breaking the LET at  order of $1/a^2$, where 
$a=\frac{m_\rho^2}{g_\rho^2 f_\pi^2}=1.84$ 
and $m_\rho$ 
is the empirical mass of the $\rho$-meson. This conclusion is based on the assumption that VMD is valid.   
\vspace{0.5cm}

 \includegraphics[height=3.5cm,angle=0]{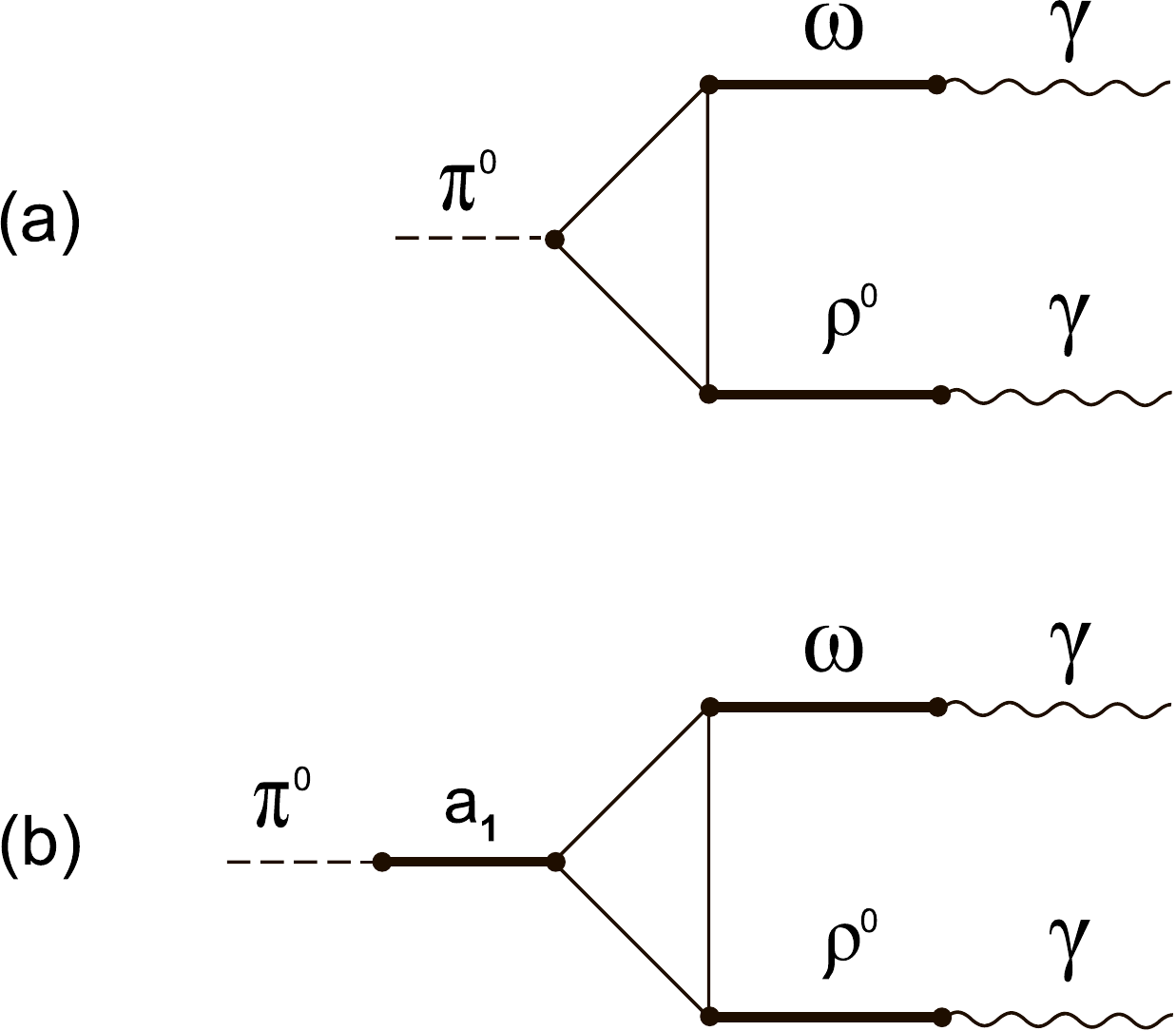} \hspace{0.3cm} {\includegraphics[height=3.5cm,angle=0]{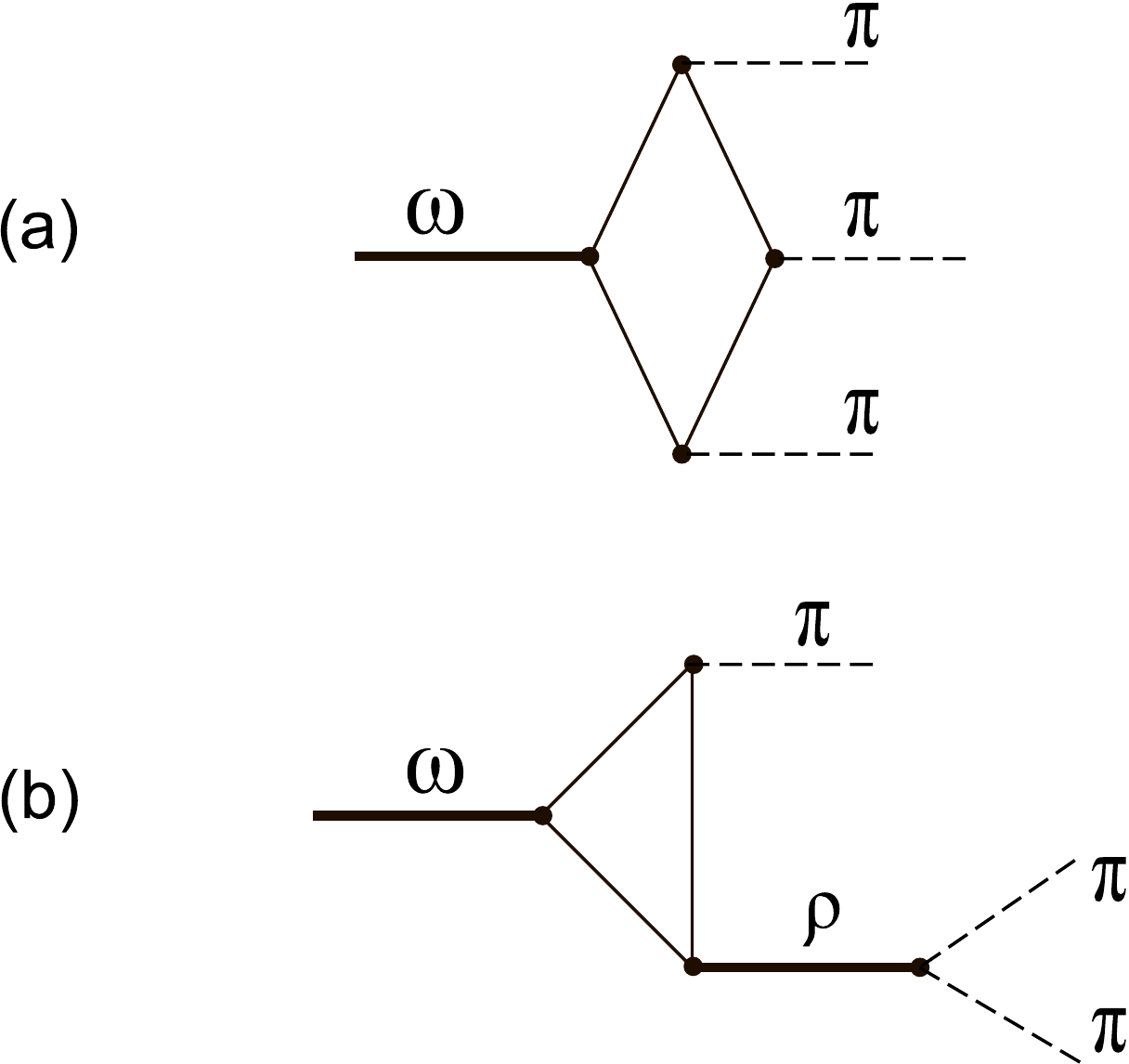} \hspace{0.3cm}  {\includegraphics[height=2.cm,angle=0]{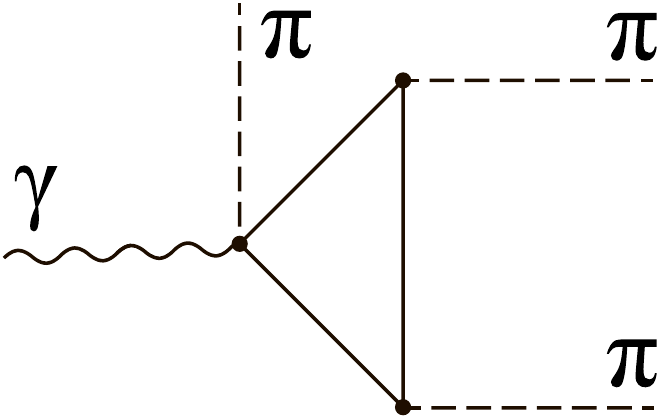}

{\small Fig. 1. Left (a) and (b): the two graphs describing the $\pi^0\to\gamma\gamma$ decay in the NJL model, (b) for $\pi a_1$-mixing effects on the pion line; Middle:  quark loop contributions to $\omega\to 3\pi$ decay, (a)  full set of possible diagrams without and with 1, 2, and 3 $\pi a_1$-mixing effects on the pion line (not drawn);  (b) $\rho$ exchange diagrams without and with $\pi a_1$ transitions; 
Right: contribution to $\gamma\to 3\pi$ decay  due to covariant $\pi a_1$ diagonalization, see (\ref{cov}), with  pion lines subject to $\pi a_1$-mixing.} 
\vspace{0.5cm}

Let us recall and  complement the calculations made in \cite{Wakamatsu89}. The diagrams contributing to the $\omega\to 3\pi$ decay are shown in Fig. 1, middle, where we have additionally taken into account the box diagram with three $\pi a_1$-transitions in (a) as well as the contribution of the $\omega\rho (a_1\to\pi )$ vertex in the $\rho$-exchange graph (b),  both  neglected in \cite{Wakamatsu89}. The corresponding amplitude is  given by
\begin{equation}   
\label{om3pi} 
A_{\omega\to 3\pi}=-\frac{N_c g_\rho}{4\pi^2 f_\pi^3} e_{\mu\nu\alpha\beta} \epsilon^\mu (q) p_0^\nu p_+^\alpha p_-^\beta 
F_{\omega\to 3\pi},
\end{equation}
where $p_0, p_+, p_-$ are the momenta of the pions, $\epsilon^\mu (q)$  the polarization of the $\omega$-meson with momentum $q$, and the form factor $F_{\omega\to 3\pi}$ is found to be
\begin{eqnarray}
\label{ff}
F_{\omega\to 3\pi}&=&\left(1-\frac{3}{a}+\frac{3}{2a^2}+\frac{1}{8a^3}\right) 
+ \left(1-\frac{c}{2a}\right)\sum_{k=0,+,-} \frac{g_\rho^2 f_\pi^2}{m_\rho^2-(q-p_k)^2}.
\end{eqnarray}
In the first parentheses, the box diagrams without, with one, two, and three $\pi a_1$-transitions are given correspondingly. The last term represents the contribution of $\rho$-exchange graphs, where $c$ controls the magnitude of an arbitrary local part of the anomalous AVV-quark-triangle. 
In the low-energy limit, the sum yields $3/a$, as one neglects the dependence on momenta in (\ref{ff}), leading to full cancellation among the terms of order $1/a$,  as well-known  \cite{Wakamatsu89}. The ST $c$ contributes at order $1/a^2$. For $c=0$ we reproduce the $\pi a_1$-mixing effect found in \cite{Wakamatsu89} to this order. 
Had $c$ been used instead to cancel the $\pi a_1$-mixing effect, as  $c=1+1/(12a)$, a too low width $\Gamma (\omega\to\pi^+\pi^0\pi^-)=3.2\,\mbox{MeV}$ would be obtained as compared to experiment  $\Gamma (\omega\to\pi^+\pi^0\pi^-)=7.57\pm 0.13 \,\mbox{MeV}$.
Furthermore the value $c=0$ is also required following \cite{Cohen89}, where the chiral Ward identities (WI) for  $\gamma\to 3\pi$  imply that both the chiral triangle and the box anomaly contribute as
\begin{equation}
A_{\gamma\to 3\pi}^{tot}=\frac{3}{2}A^{AVV}-\frac{1}{2}A^{VAAA}, 
\end{equation}   
where $A_{\gamma\to 3\pi}^{tot}$, $A^{AVV}$ and $A^{VAAA}$ are, respectively, the total $\gamma\pi\pi\pi$  amplitude, the $\gamma\to\omega\to\pi\rho\to\pi\pi\pi$ process and the point $\gamma\to\omega\to\pi\pi\pi$ amplitude. This result is consistent with both the chiral WI  and with the KSFR relation \cite{Kaw66,Riaz66}, 
which arises in the NJL model at $a=2$. One sees from eq. (\ref{ff}) that, if one neglects the terms of order $1/a^2$ and higher in the box contribution and puts $c=0$ in the $\rho$-exchange term, the amplitude $A^{VAAA}$  has a factor $(1-3/a)=-1/2$, and the $A^{AVV}$ amplitude has a factor $(1-c/(2a)) 3/a=3/2$, as is required by the chiral WI. On the other hand, if $c$ is chosen to cancel $\pi a_1$-mixing effects, these amplitudes contribute with  relative weights  $-7/64$ and $71/64$, respectively. 
Therefore the ST $c$ cannot be used to resolve the $\pi a_1$-mixing puzzle,  the chiral WI require $c=0$. This pattern has been considered in \cite{Schechter84,Kaiser90,Wakamatsu89}, and reproduces well the phenomenological value of the width. That allows us to conclude, following \cite{Wakamatsu89}, that if the VMD is a valid theoretical hypothesis, the $\gamma\to\omega\to 3\pi$ amplitude contains contributions due to $\pi a_1$-mixing that violate the LET (\ref{LET})       
\begin{equation}
\label{g3pi}
A_{\gamma\to 3\pi}=-F^{3\pi} e_{\mu\nu\alpha\beta} \epsilon^\mu (q) p_0^\nu p_+^\alpha p_-^\beta , 
\end{equation}
\begin{equation}
\label{3pi}
F^{3\pi}=\frac{N_c e}{12\pi^2 f_\pi^3}\left(1+\frac{3}{2a^2}+\frac{1}{8a^3}\right)\neq \frac{N_c e}{12\pi^2 f_\pi^3}.
\end{equation}
In the following we will show that it is possible to combine the phenomenologically successful value $c=0$ with a full cancellation of $\pi a_1$-mixing effects  within the NJL approach by taking into account the  anomalous AAA triangle shown in the right side of Fig.1, which occurs as result of (\ref{cov}) 
\begin{eqnarray}
A&=&\frac{N_c e}{4a^3f_\pi^3}  
\left\{p_-^\sigma [J_{\mu\nu\sigma} (p_0,p_-)-J_{\mu\sigma\nu} (p_-,p_0)] \right. \nonumber \\
&& \!\!\!\! + \left. p_+^\sigma [J_{\mu\nu\sigma} (p_0,p_+)-J_{\mu\sigma\nu} (p_+,p_0)] \right\} \epsilon^\mu (q)p_0^\nu . 
\end{eqnarray}
The low energy expansion of the loop integral $J_{\mu\nu\sigma}$ starts from a linear term
\begin{equation}
\label{exp}
J_{\mu\nu\sigma} (p_0,p_-)=\frac{1}{24\pi^2} e_{\mu\nu\sigma\rho}\left(p_0-p_-  -3 \upsilon \right)^\rho +\ldots
\end{equation}
Owing to the shift ambiguity related to the formal linear divergence of this integral, the result depends on the undetermined 4-vector $\upsilon_\rho$, 
\begin{equation}
\label{amp}
A=-\frac{N_c e}{4\pi^2 f_\pi^3} e_{\mu\nu\sigma\rho}\epsilon^\mu (q)p_0^\nu (p_++p_-)^\sigma \left(\frac{\upsilon^\rho}{4a^3}\right)
\end{equation}
This is the complete result for this triangle diagram. 
The  4-vector $\upsilon_\rho$  is represented as linear combination of the independent momenta of the process, $\upsilon_\mu= b_1 q_\mu + b_2 (p_+ -p_-)_\mu + b_3 (p_+ + p_-)_\mu$, but only the second term survives in (\ref{amp}). Thus, the graph shown on  Fig.1 right gives an additional contribution $\Delta F^{3\pi}$ to the form factor $F^{3\pi}$   
\begin{equation}   
\label{g3pinew} 
\Delta F^{3\pi}=\frac{N_c e}{12\pi^2 f_\pi^3} \left(\frac{-3b_2}{2a^3}\right),
\end{equation}
where $b_2$ is dimensionless and as yet undetermined.  This constitutes a further example in which an arbitrary regularization dependent parameter should be fixed by the physical requirements \cite{Jackiw00, Baeta01,Batista18}.  The AAA amplitude would be zero had it been regularized in advance by any regularization that sets ST to zero. For a detailed discussion of this and further anomalous vertices appearing in the present calculation we refer to \cite{okh:2020}.
To fix $b_2$  we use the LET (\ref{LET}); requiring that the unwanted terms in (\ref{3pi}) vanish we find that $b_2=a+\frac{1}{12}=1.92$.
Thus, the solution of the $\pi a_1$-mixing problem in the $\gamma\to 3\pi$ amplitude can be associated with the ST of the anomalous non-VMD diagram shown on the right of Fig. 1.




\end{document}